\documentclass[%
 reprint,
nofootinbib,
 amsmath,amssymb,
 aps,
 onecolumn
]{revtex4-2}

\usepackage{amssymb,amsmath,amsfonts,amsbsy,graphicx}
\usepackage{color}
\usepackage{psfrag}
\usepackage{stackrel}

\usepackage[colorlinks=true, allcolors=blue]{hyperref}

\newcommand\be{\begin{equation}}
\newcommand\ee{\end{equation}}
\newcommand\ba{\begin{eqnarray}}
\newcommand\ea{\end{eqnarray}}\newcommand\eq{\begin{equation}}           
\newcommand\en{\end{equation}}

 \newcount\colveccount
\newcommand*\colvec[1]{
        \global\colveccount#1
        \begin{pmatrix}
        \colvecnext
}
\def\colvecnext#1{
        #1
        \global\advance\colveccount-1
        \ifnum\colveccount>0
                \\
                \expandafter\colvecnext
        \else
                \end{pmatrix}
        \fi
}
\input epsf

\usepackage{ulem}
\usepackage{amssymb}
\usepackage{amsmath}
\usepackage{bm}
\usepackage{graphicx}
\usepackage{color}
\usepackage{subfig}
\usepackage{caption}

\usepackage{tikz}
\usetikzlibrary{positioning}
\usepackage{siunitx}
\usetikzlibrary{intersections}

\def\gsim{\;\rlap{\lower 2.5pt
 \hbox{$\sim$}}\raise 1.5pt\hbox{$>$}\;}
\def\lsim{\;\rlap{\lower 2.5pt
 \hbox{$\sim$}}\raise 1.5pt\hbox{$<$}\;}

\usepackage{graphicx}
\usepackage{url}
\usepackage{amssymb,amsmath}
\usepackage{amsbsy}

\newcommand{\mnras}{MNRAS}

\usepackage{graphicx}
\usepackage{url}
\usepackage{amssymb,amsmath}
\usepackage{amsbsy}

\usepackage{svg}
\newcommand{\orcid}[1]{$\,$\href{https://orcid.org/#1}{\includesvg[width=10pt]{orcid}}}

\begin{document}
\title{
 Boosting the 21 cm forest signals by the clumpy substructures
}
\author{Kenji Kadota$^{1,2}$,
  Pablo Villanueva-Domingo$^3$,
  Kiyotomo Ichiki$^{4,5}$, Kenji Hasegawa$^{4}$, Genki Naruse$^{4}$\\
  {\small  $^1$School of Fundamental Physics and Mathematical Sciences,
  Hangzhou Institute for Advanced Study,\\
  University of Chinese Academy of Sciences (HIAS-UCAS), Hangzhou 310024, China\\
  $^2$International Centre for Theoretical Physics Asia-Pacific (ICTP-AP), Beijing/Hangzhou, China\\
  $^3$
  Computer Vision Center - Universitat Aut\`onoma de Barcelona (UAB), 08193 Bellaterra, Barcelona, Spain
  \\
  $^4$
  Graduate School of Science, Division of Particle and Astrophysical Science, Nagoya University, Nagoya, 464-8602, Japan
$^5$Kobayashi-Maskawa Institute for the Origin of Particles and the Universe, Nagoya University, Nagoya, 464-8602, Japan
  }
}

\begin{abstract}
  We study the contribution of subhalos to the 21 cm forest signal.
  The halos can host the substructures and including the effects of those small scale clumps can potentially boost the 21 cm optical depth in favor of detecting the 21 cm forest signals. 
  We estimate the boost factor representing the ratio of the optical depth due to the subhalo contribution and that due to the host halo alone (without subhalos).
  Even though the optical depth boost factor is negligible for a small host halo with the mass of order $10^5 M_{\odot}$, the subhalo contribution can enhance the optical depth by an order of magnitude for a host halo of order $10^7 M_{\odot}$. The resultant 21 cm absorption line abundance which is obtained by integrating over the halo mass range relevant for the 21 cm forest signal can be enhanced by up to of order $10\%$ due to the substructures.
  The larger boost factor for a larger host halo would be of particular interest for the 21 cm forest detection because the the contribution of the larger host halos to the 21 cm forest signals is smaller due to their higher temperature and less abundance than the smaller host halos. The subhalos hence can well help the larger host halos more important for the signal estimation which, without considering the subhalos, may not give appreciable contribution to 21 cm forest signals. 
\end{abstract}

\maketitle   

\setcounter{footnote}{0} 
\setcounter{page}{1}\setcounter{section}{0} \setcounter{subsection}{0}
\setcounter{subsubsection}{0}

\section{Introduction}


The 21 cm forest can provide promising probes on the neutral hydrogen (HI) gas content of the Universe before the reionization epoch \cite{1959ApJ...129..525F,2002ApJ...577...22C,2002ApJ...579....1F,2006MNRAS.370.1867F, Chabanier:2018rga,2015aska.confE...6C}.
The 21 cm forest refers to the absorption spectra along the line of sight to the radio bright sources, in an analogous way to the Lyman-$\alpha$ forest. In contrast to too large an optical depth at the Lyman-$\alpha$ resonance frequency at the pre-ionization epoch (Lyman-$\alpha$ is hence a better probe on the highly ionized gas at the post-ionization epoch), a much weaker 21 cm hyperfine transition of the neutral hydrogen can provide us a unique probe to study the pre-reionization epoch whose detailed evolution still remains largely unknown. 
A promising source of such an absorption system is the so-called minihalo: star-less gas cloud which is too small to host the stars. The virial temperature is below atomic hydrogen line cooling threshold, so that the neutral hydrogen can stay neutral without being ionized in the absence of the efficient heating sources well before the reionization epoch.
Minihalos filled with the neutral hydrogen can resonantly absorb the radiation which redsfhits to the local 21 cm transition frequency.

In the standard $\Lambda$CDM cosmology, structures form hierarchically. The small halos form first and they can merge and accrete to form larger halos. If those small halos survive in a larger host halo, those small scale clumpy structures inside a host halo can potentially imprint the observable signatures.
Because minihalos are also expected to host the subhalos, 
the 21 cm forest can be a promising way to study the gas properties of the subhalos in a minihalo. This would be of particular interest because the HI content of the subhalos and their evolution at the pre-ionization epoch are not well constrained. 
While the 21 cm forest estimation due to minihalos has been discussed without considering such clumpiness in a minihalo so far, our goal is to point out the possibility that the radio absorption spectra can well be enhanced due to such substructures. 

The signal boosting due to the small scale structures compared with those without considering these clumpy structures is often characterized by a boost factor (also called a clumping factor). For instance, the boost factor for the hydrogen recombination rate characterized by the (ionized hydrogen) gas density squared 
has been extensively discussed, which can potentially affect the reionization history of the Universe \cite{Finlator:2012gr,Haiman:2000pd, Bianco:2021ccu,Mao:2019vob,2006PhR...433..181F, 1999ApJ...514..648M, 1997ApJ...486..581G}. 
 Another common place where the signal boosting by the clumpy substructures has been discussed is the dark matter annihilation which is proportional to the dark matter density squared \cite{Ullio:2002pj,Taylor:2002zd} \footnote{Note the dark matter decay signals, rather than the annihilation, are linearly proportional to the dark matter density rather than the density squared, and the boost factor does not show up (because the over-dense and under-dense regions just linearly cancel out).}.
Our 21 cm forest signals on the other hand linearly depend on the gas density and the overdense regions can contribute to the additional deep absorption spectra.

We calculate the boost factor of the optical depth along the line of sight to a high-$z$ radio source defined as the ratio of the subhalo optical depth in a given host halo to that of a host halo without including the subhalo contribution. While dedicated numerical simulations are required to study the gas properties in the subhalos, as the first attempt to study the potential significance of subhalos to the 21 cm forest, the aim of our paper is to provide a simple analytical estimation on how much boosting of the 21 cm forest signals can be expected due to the subhalos.
We show that the subhalos can boost the 21 cm optical depth by an order of magnitude for a large minihalo with mass $\sim 10^7 M_{\odot}$ and the abundance of the spectrum absorption lines can increase by up to of order $10 \%$. Even though our estimation suffers from the large uncertainties in the gas properties at the pre-ionization epoch, such as the tidal disruption and gas heating which may diminish the absorption abundance enhancement to a few per cent level, our study can serve as a good motivation for further exploration of 21 cm forest as a tool to study the small scale structures in the pre-ionization epoch.

We mention that the boosting of the 21 cm forest signals has also been discussed in the scenarios beyond the simple $\Lambda$CDM cosmology, such as those with the enhanced primordial isocurvature density fluctuations due to the axion-like particle dark matter or PBH dark matter \cite{Shimabukuro:2019gzu,Shimabukuro:2020tbs,Kawasaki:2020tbo,Villanueva-Domingo:2021cgh}. The 21 cm forest signal has been shown to give a promising probe on the nature of those dark matter scenarios. Our paper is different from those previous papers discussing the enhancement of the minihalo abundance due to the enhancement of the structure formation in the presence of the additional power in the dark matter fluctuations. Those papers discuss the scenarios where 21 cm forest signals are boosted because the minihalo formation epoch occurs earlier and the minihalo abundance is bigger compared with the conventional $\Lambda$CDM cosmology scenarios. Our study instead deals with more generic scenarios in that our focus is to study the boosting of 21 cm signals due to the subhalos inside the minihalos in the standard $\Lambda$CDM cosmology.

Our paper is structured as follows. Section \ref{sec:review} outlines the analytical formalism to estimate the 21 cm forest signals from the minihalos. Based on the standard calculations reviewed in Section \ref{sec:review}, Section \ref{sec:minihalo} presents our modeling of the minihalo's subhalo contribution to the optical depth for the 21 cm absorption along the line of sight towards a radio source. Section \ref{sec:results} presents the quantitative results based on our formalisms and we also estimate how much enhancement in the absorption line abundance can be realized due to the subhalos. Section \ref{sec:disc} is devoted to the discussion and conclusion.

\section{Formalism}
\subsection{21 cm absorption from a minihalo}
\label{sec:review}
The minihalos are dense gas clouds which are virialized but too small to form stars. Such halos are hence expected to be the reservoirs of neutral hydrogen and the minihalos can lead to the system of 21 cm absorption lines, in analogy to the Lyman-$\alpha$ forest, when the flux from a radio source such as a radio-loud quasar goes through the HI rich minihalos.
We here outline the relevant characteristics of a minihalo and its optical depth for 21 cm absorption following Refs. \cite{2002ApJ...579....1F,2001PhR...349..125B,2006MNRAS.370.1867F,2006PhR...433..181F, Villanueva-Domingo:2021cgh}.
\\
One first needs to specify the minihalo profiles (dark matter and gas profiles) to estimate the 21 cm optical depth. We assume the Navarro, Frenk $\&$ White (NFW) profile \cite{1997ApJ...490..493N,2000ApJ...540...39A} for the dark matter distribution characterized by the concentration parameter $c\equiv R_{vir}/R_s$ 
\ba
\rho_{\rm NFW}(R)=\frac{c^3}{3f(c)}
\frac{\Delta_c \rho_c(z)}{\frac{R}{R_s} \left( 1+ \frac{R}{R_s}\right)^2 }
\label{eq:nfw}
\ea
where $f(x)=\ln(1+x)-x/(1+x)$.
$R_s$ is the scale radius and the virial radius is \cite{2001PhR...349..125B}
\ba
  R_{{\rm vir}}=0.784\bigg(\frac{M}{10^{8}h^{-1}M_{\odot}}\bigg)^{1/3}\bigg[\frac{\Omega_{m}}{\Omega_{m}^{z}}
  \frac{\Delta_{c}} {18\pi^{2}}\bigg]^{-1/3} \bigg(\frac{1+{\it z}}{10}\bigg)^{-1}h^{-1}[{\rm kpc}]
\ea
where $M$ is the halo mass, $\Delta_{c}=18\pi ^{2}+82d-39d^{2}$ is the over-density of a virialized halo collapsing at the redshift $z$, $d=\Omega_{m}^{z}-1$ and
$\Omega_{m}^{z}=\Omega_{m}(1+z)^{3}/(\Omega_{m}(1+z)^{3}+\Omega_{\Lambda})$. For the concentration parameter $c(M,z)$, we adopt the fitting formula of Ref. \cite{Ishiyama:2020vao} based on the Uchuu simulations \cite{uchuuwebpage} for its wide coverage of the mass and redshift relevant for our study. 
We assume that the corresponding gas profile is in hydrostatic equilibrium and isothermal (settled at the virial temperature), for which the gas density profile can be derived analytically \cite{1998ApJ...497..555M,2011MNRAS.410.2025X,Suto:1998xs}
\ba
 \ln \rho_{{\rm g}}(R)=\ln \rho_{{\rm g0}}-\frac{\mu m_{{\rm p}}}{2k_{{\rm B}}T_{{\rm vir}}}[v_{{\rm esc}}^{2}(0)-v_{{\rm esc}}^{2}(R)],
\label{eq:gas_profile1}
\ea
where the virial temperature is defined as $T_{vir}=(\mu /2k_B) (GM/R_{vir})$ ($\mu$ is the mean molecular weight and $\mu \sim 1.2$ for the neutral primordial gas) \cite{2001PhR...349..125B}.
$v_{{\rm esc}}(R)$ is the escape velocity
\begin{equation}
  v_{{\rm esc}}^{2}(R) = 2\int_{R}^{\infty}\frac{GM(R^{'})}{R^{'2}}dr^{'}=2V_{c}^{2}\frac{f(cx)+cx/(1+cx)}{xf(c)}~,
\label{eq:esc_velocity}
\end{equation}
where $x\equiv R/R_{{\rm vir}}$ and $V_{c}$ is the circular velocity
\ba
  V_{c} =  \sqrt{\frac{GM}{R_{{\rm vir}}}}=23.4\bigg(\frac{M}{10^{8}h^{-1}M_{\odot}}\bigg)^{1/3}
  \bigg[\frac{\Omega_{m}}{\Omega_{m}^{z}}\frac{\Delta_{c}}{18\pi^{2}}\bigg]^{1/6}  \bigg(\frac{1+z}{10}\bigg)^{1/2}[{\rm km/s}].
  \label{eq:cir_velocity}
\ea
The central density $\rho_{g0}$ is normalized by demanding that the total baryonic mass fraction in a halo to be $\Omega_{b}/\Omega_{m}$,  leading to
\ba
  \rho_{g0}(z) =
  \frac{(\Delta_c/3)c^{3}e^{A}}{\int_{0}^{c}(1+t)^{A/t}t^{2}dt} \left(\frac{\Omega_b}{\Omega_m}\right) \rho_{c}(z)~,
\ea
with $A=3c/f(c)$. Once the profile for the minihalo is specified, one can estimate the optical depth for the 21 cm absorption along the line of sight at an impact parameter $\alpha$ from the halo center \cite{2002ApJ...579....1F,2001PhR...349..125B,Meiksin:2011gx,Xu:2010us}
\ba
  \tau(M,\alpha)=\frac{3h_{{\rm p}}c^{3}A_{10}}{32\pi k_{{\rm B}}\nu_{21}^{2}}
  \int_{-l_{{\rm max}}(\alpha)}^{l_{{\rm max}}(\alpha)}dl\frac{n_{{\rm H{\sc I}}}(R)}{T_{{\rm S}}(R)\sqrt{\pi}b} \exp\bigg(-\frac{v^{2}}{b^{2}}\bigg),
  \label{taueq1}
  \ea
  where $l$ is the integration along the line of sight with $l=\sqrt{R^2-\alpha^2}$, $l_{\mathrm{max}}=\sqrt{R^2_{vir}-\alpha^2}$ and $A_{10}$ is the Einstein coefficient for the spontaneous transition. The exponential factor represents the Doppler broadening with $v=c(\nu-\nu_{21})/\nu_{21}$ and the velocity dispersion $b^2=2kT_{vir}/m_p$ ($\nu_{21}=1420$MHz$=\nu/(1+z)$ is the rest frame hyperfine transition frequency) \cite{2002ApJ...579....1F,1979rpa..book.....R}. 
$T_S$ is the spin temperature in a minihalo \cite{2005ApJ...622.1356Z,2006MNRAS.370.1867F,2014PhRvD..90h3003S}, and it approaches the virial temperature in the inner halo region and it becomes the CMB temperature at the outer region (as expected because the gas density is big (small) in the inner (outer) region of a halo). One can then estimate the abundance of absorption lines as  
\ba
\frac{dN(>\tau)}{dz}
=
\frac{dr}{dz} \int^{M_{max}}_{M_{min}} dM_h \frac{dN_{h}}{dM_h}
\pi R^2_{ \tau }(M,\tau)
\label{dndz}
\ea
where $dr/dz=c/H(z)$ is the comoving line element, $dN_h/dM_h$ is the halo mass function for which we adopt Ref. \cite{1999MNRAS.308..119S,Sheth:1999su} and we consider the geometric cross section $\pi R^2_{\tau }$ such that the optical depth exceeds a given value of $\tau$ within an impact parameter $ R_{\tau }$.
For the mass integration, we set the lower limit as the mass characterized by the Jeans scale of the intergalactic medium (IGM) $\lambda_J=\sqrt{5 \pi k_B T_{IGM} / 3G \mu m_p \rho_m(z)}$
\ba
M_{min}= \frac{4 \pi \rho_m(z)}{3} \left( \frac{\lambda_J}{2}  \right)^3
\simeq
4\times 10^4 M_{\odot} h^{-1} \left( \frac{T_{IGM}}{10K} \right)^{3/2}
\left( \frac{10}{1+z}  \right)^{3/2}
\label{mmin}
\ea
The value of IGM temperature at a high redshift is not well constrained. The current observation bounds allow as low as $T\sim 10$K around $z\sim 8$ \cite{HERA:2021noe,Greig:2020suk,2015ApJ...809...62P}. The temperature at a higher redshift can be lower if the heating is still not efficient. The 21 cm forest signal amplitude is sensitive to the lower limit of this mass integration or equivalently the value of IGM temperature as discussed in the literature \cite{2002ApJ...579....1F,2006MNRAS.370.1867F,1998MNRAS.296...44G}. We present our discussions adopting the adiabatic cooling of IGM \cite{2001PhR...349..125B,2003tsra.symp...31M} unless stated otherwise, and we illustrate how sensitive our results are on the change of the IGM temperature at the end of Section \ref{sec:results}. We set the integration mass upper limit to be the virial mass corresponding to $T_{vir}\sim 10^4 K$ to prevent the star formation via the atomic cooling
\ba
M_{max}\simeq 4\times 10^7 M_{\odot} h^{-1} \left( \frac{T_{vir}}{10^4 K} \right)^{3/2}
\left(
\frac{\Omega_m}{\Omega^z_m}
\frac{\Delta_c}{18\pi^2}
\right)^{-1/2}
\left(
\frac{1+z}{10}
\right)^{-3/2}
\ea

Having outlined how to estimate the 21 cm forest signals for a given halo, we now discuss how the subhalo contribution to the 21 cm forest signal can be estimated in the following section. 


\subsection{Boosting 21 cm absorption by subhalo contributions}
\label{sec:minihalo}

\begin{figure}[h]
\begin{tikzpicture}[baseline]

\draw (0,0) circle (3cm);
\draw (-3,0) -- (3,0);
\draw (0,-3.)--(0,3);

\draw (0.9,1.)--(2.3,1.);
\draw[dashed] (0.9,-3.1)--(0.9,3.1) node[above] {$x_{-}$};
\draw[dashed] (2.3,-3.1)--(2.3,3.1) node[above] {$x_{+}$};
\draw[line width=0.5mm ] (1.6,-3.7)--(1.6,3.7) node[above] {$LOS$ at $\alpha_{h}$};


\node at (1.95,1.15) {$r_{vir}$};
\node at (-1.5,-0.16) {$R_{vir}$};

\end{tikzpicture}
        \caption{The number of subhalos encountered by a line of sight $N^{LOS}_{s}$ can be estimated by considering the overlapping volume of the host halo and the column of radius $r_{vir}$ around the line of sight at an impact parameter $\alpha_{h}$.}
\label{fig:sphere1}
\end{figure}
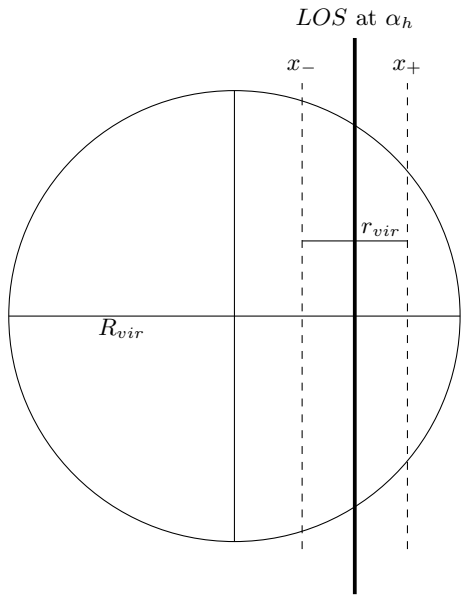

The previous section illustrates the simple estimation for the optical depth for isolated (host) halos by assuming the smooth density profile (e.g. the NFW profile) without considering the clumpy substructures inside them.
To study the contribution of those subhalos inside a host halo, we first estimate the number of subhalos $N^{LOS}_{s}$ encountered by a line of sight (LOS) at a given impact parameter $\alpha_{h}$ from a host halo center. For this purpose, we consider the subhalo mass function $dn_{s}/dm_{s}(M_{h})$ (the differential number of subhalos in a subhalo mass interval for a given host mass $M_{h}$). The differential number of subhalos along the LOS is given by the subhalo mass function multiplied by the overlapping volume between the host halo and a cylinder around the LOS with a radius given by the subhalo virial radius $r_{vir}(m_{s})$ as illustrated in Fig. \ref{fig:sphere1}, weighted by a normalized subhalo distribution $w(r)$, which reads
\ba
\frac{d N^{LOS}_{s}}{dm_{s}}(M_{h},\alpha_{h},m_{s})
=\frac{d n_{s}(M_{h})}{dm_{s}}
4 \int_{x_-}^{x_+} dx
\int_0 ^{y_+} dy \int_{0}^{z_+} dz 
w(\sqrt{x^2+y^2+z^2})  
\label{eq:overlap}
\ea
where $x_{+}=min(\alpha_h+r_{vir},R_{vir}), x_{-}=max(\alpha_h-r_{vir},-R_{vir}), y_+=min(\sqrt{r_{vir}^2-(x-\alpha_h)^2},\sqrt{R_{vir}^2-x^2}), z_+=\sqrt{R_{vir}^2-x^2-y^2}$. $R$ and $r$ represent the radius of the host halo and subhalo respectively, i.e., $R_{vir}$ ($r_{vir}$) is the virial radius of the host halo (subhalo), with the subscript $h$ ($s$) representing the host halo (subhalo). Our following discussions assume $\alpha_h\geq 0, R_{vir} -r_{vir}\geq 0$ without the loss of generality. The HI-hosting subhalo mass function $dn_{s}/dm_{s}$ for a given halo mass $M_{h}$ at a high redshift is not well constrained and there are relevant issues under activate debate, such as the spatial distribution of subhalos in a host halo and the subhalo survival under the influence of tidal disruption \cite{Spinelli:2019smg, Villaescusa-Navarro:2018vsg,Han:2015pua,Springel:2008cc,Stref:2016uzb,Kelley:2018pdy,Hayashi:2002qv} \footnote{See for instance Refs. \cite{Spinelli:2019smg, Villaescusa-Navarro:2018vsg,2018MNRAS.478..548S,2021MNRAS.506.1507C,2013MNRAS.430.2427R} for the modeling of the gas profile in a halo at a low (post-reionization) redshift.}. We for concreteness use the subhalo mass function of Ref. \cite{Hiroshima:2018kfv,Ando:2019xlm} which was calibrated to the compilation of N-body simulation results covering a wide range of mass (from galaxy clusters down to the Earth mass $\sim 10^{-6} M_{\odot}$) (the explicit form is also given in the appendix).
We also introduce the subhalo distribution weighting function $w(R)$ defined as 
\ba
w(R)=\frac{s(R) }{\int s(R) d^3 R},
\label{eq:weight1}
\ea
where $\int d^3 R =\int _0^{R_{vir}} dR 4 \pi R^2$ represents the integration over the host halo.
For instance, if we assume the uniformly distributed subhalos inside a host halo, $s(R)=const$.
If we assume the subhalos follow the underlying host halo dark matter mass density (for instance the NFW profile from Eq. \ref{eq:nfw} for the host halo density $\rho_{h}$)
\ba
w(R)=\frac{\rho_{h}(R)}{ \int d^3 R \rho_{h}(R)}=\frac{\rho_{h}(R)}{ M_{h}}
\label{eq:weight2}
\ea
In estimating the subhalo contribution to the optical depth, we make a simplification by treating a subhalo as a point-like object inside a host halo and consider the optical depth averaged over a geometric cross section of a single subhalo $\pi r_{vir}^2(m_{s})$
\ba
\tau_{ave}(m_{s})=
\frac{2\pi \int _{0}^{r_{vir}} d \alpha \alpha \tau(m_{s},\alpha)}
     {
2\pi \int_0^{r_{vir}} \alpha  d\alpha 
     } 
=\frac{2\pi \int d\alpha  \alpha \tau(m_{s},\alpha)} {\pi r_{vir}^2 } 
\ea
where we adopt Eq. \ref{taueq1} for $\tau(m_{s},\alpha)$.
The HI distribution inside a subhalo at a high redshift is not well known and we simply use the isothermal gas profile assuming the virial temperature of the subhalo following the formalism presented in the last section. We can then integrate over the subhalo mass within a single host halo
\ba
 \tau^{s}  (M_{h},\alpha_{h})
=\int   dm_{s}\frac{d N^{LOS}_{s}}{dm_{s}}(M_{h},\alpha_{h},m_{s})\times \tau_{ave}(m_{s})
\ea
We need to specify the lower mass in integrating over the subhalo mass. We for concreteness use the Jeans mass at the observation redshift (Eq. \ref{mmin}). We mention that, at the redshift where there can exist the X-ray/UV radiation sources, the small mass (sub)halos do not host abundant HI because they are ionized due to the lack of self-shielding against ionizing radiation. We focus on a high redshift where there are still not yet sufficient ionizing sources, and the lower mass cut for the (sub)halos was introduced from the Jeans mass scale rather than from the requirement of self-shielding. A better choice could be the Jeans mass at the epoch of the subhalo's accretion to the host halo. The observation redshift of our interest is $z_{obs}\gtrsim 10$ corresponding to the age of the Universe $\lesssim 0.5$ Gyr. The epoch of subhalo's infall to the host halo and the observation epoch hence would not differ appreciably in the parameter range of our interest (the difference would be at most of order ${\cal O}(0.1)$ Gyr). We defer the discussions on the subhalo gas properties including their dynamics such as the subhalo mass loss rate inside a host halo to our forthcoming simulation paper.
The upper bound in the subhalo mass integration is incorporated in the subhalo mass function because it has an exponential cutoff when the subhalo mass becomes close to the host halo mass \cite{Hiroshima:2018kfv,Ando:2019xlm}. The total optical depth of a host halo including subhalos can be estimated as the sum of the host and subhalo contributions 
\ba
 \tau (M_{h},\alpha_{h})
=  \tau^{h}  (M_{h},\alpha_{h})
+ \tau^{s}  (M_{h},\alpha_{h})
=\left(1+B \right)    \tau^{h}
\ea
where the boost factor of the optical depth is defined as the ratio of host and subhalo contributions $ B\equiv   \tau^{s}  /   \tau^{h}$.

We can now apply our analytical formalism to quantitatively estimate the subhalo contribution to the optical depth. In the following discussions, in numerically performing the triple integration Eq. \ref{eq:overlap} by considering the intersection of the host halo sphere and the cylinder centered at the line of sight, we make the following simplifications. The boundary of a host halo may not be actually clear and model dependent (we adopted the virial radius as the halo size) and we conservatively consider the parameter range $\alpha_h+r_{vir}<R_{vir}$ (the cylinder with the radius $r_{vir}$ fits inside the host halo). The subhalos are smaller than the host halos and we further simplify our calculations by considering the cylinder with the radius $r_{vir}$ and the height $\sqrt{R_{vir}^2-(\alpha_h+r_{vir})^2}$ (we ignore the polar cap elements in the intersection and this corresponds to the choice of $z_+=\sqrt{R_{vir}^2-(\alpha_h+r_{vir})^2}$). Removing these assumptions would increase the estimated signals and hence the actual signals may be bigger than our our simplified calculations, which is hence more conservative but suffices for our goal to analytically demonstrate the potential significance of subhalos in 21 cm forest study.

All the relevant computations for the 21 cm forest have been carried out numerically in Python, based on a previous work \cite{Villanueva-Domingo:2021cgh, pablo_villanueva_domingo_2021_4707447}. We have released our implementation as the publicly available Python library HAYASHI (Halo-level AnalYsis of the Absorption Signal in HI)
\cite{pablo_villanueva_domingo_2022_7044255}.\footnote{\url{https://github.com/PabloVD/HAYASHI}}


\section{Results}

\begin{figure}[t]

\includegraphics[width=0.99\textwidth]{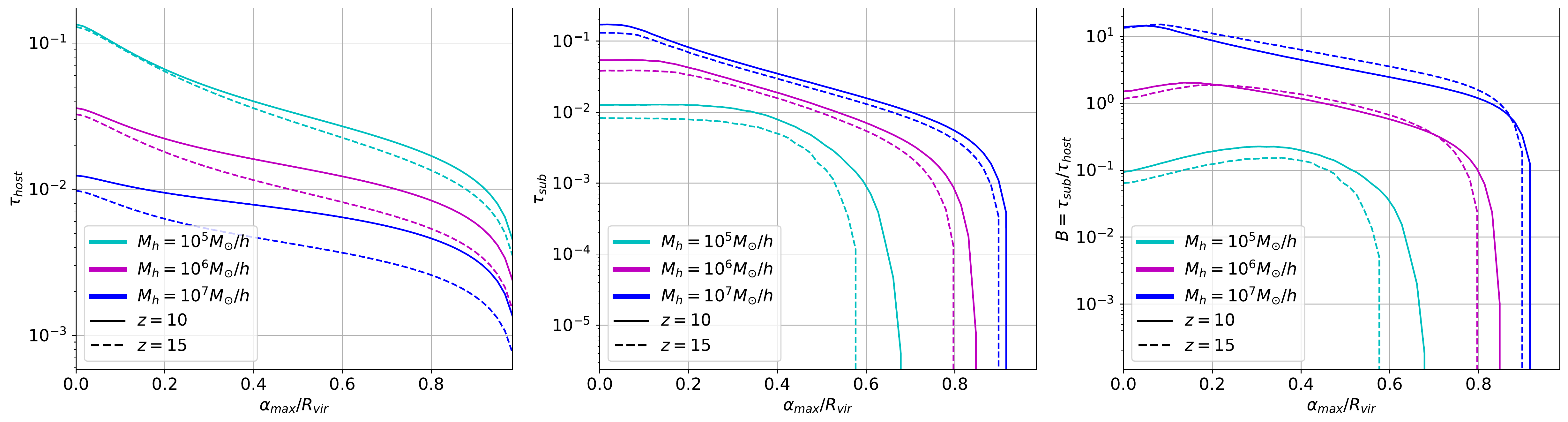}

     \caption{
       The optical depth as a function of the impact parameter normalized to the host halo virial radius at $z=10$ and $z=15$ for different host halo masses. The subscript $h$ and $s$ respectively refer to the host halo and subhalo. The total optical depth is thus given by $\tau=(1+B)\tau_{h}$, where $B=\tau_{s}/\tau_{h}$ is the boost factor, defined as the ratio between the contributions from the subhalos and the host halo alone.
       Left: The optical depth of a host halo alone without subhalo contribution.
       Middle: The subhalo contribution to the optical depth for a given host halo.
       Right: The corresponding optical depth boost factor.
   }
   \label{fig:tau}
\end{figure}

\label{sec:results}
Based on the formalisms in the last section, we can now quantitatively discuss how the subhalo contributions can affect the 21 cm forest signals with respect to the host halo contributions. The optical depth for a given host halo as a function of the impact parameter from a halo center is plotted in the left panel of Fig. \ref{fig:tau}. One can note that the optical depth is bigger in the inner halo region, where it is denser. Another notable feature is that the optical depth is bigger for a smaller halo mass because of a lower spin temperature. The subhalo contribution $\tau_{s}$ for a given host mass $M_{h}$ (middle panel) and the boost factor (right panel) are also shown in Fig. \ref{fig:tau}. The rapid decrease of the subhalo contribution for $\alpha_h\sim 1$ is because of our limiting the parameters for the subhalos to fit inside their host halos $\alpha+r_{vir}\leq R_{vir}$. The boost factor is bigger for a bigger host halo. This is expected for the following two reasons.
Firstly, the optical depth is smaller for a bigger halo because of the larger spin temperature.
Secondly, a bigger host halo possesses more abundant subhalos.
Therefore the subhalo contribution increases and host halo contribution decreases for a bigger host halo mass, so that the boost factor $B=  \tau^{s}  /   \tau^{h}$ is enhanced for a bigger host halo mass.
Even though the large boost factor for a bigger host halo is off-set by the smaller host halo mass function in estimating the 21 cm absorber abundance, it is interesting to see a large host halo whose optical depth was too small for the 21 cm observation without considering the subhalos could give a large enough optical depth due to the the subhalo boost factor.

\begin{figure}[t]

\includegraphics[width=0.99\textwidth]{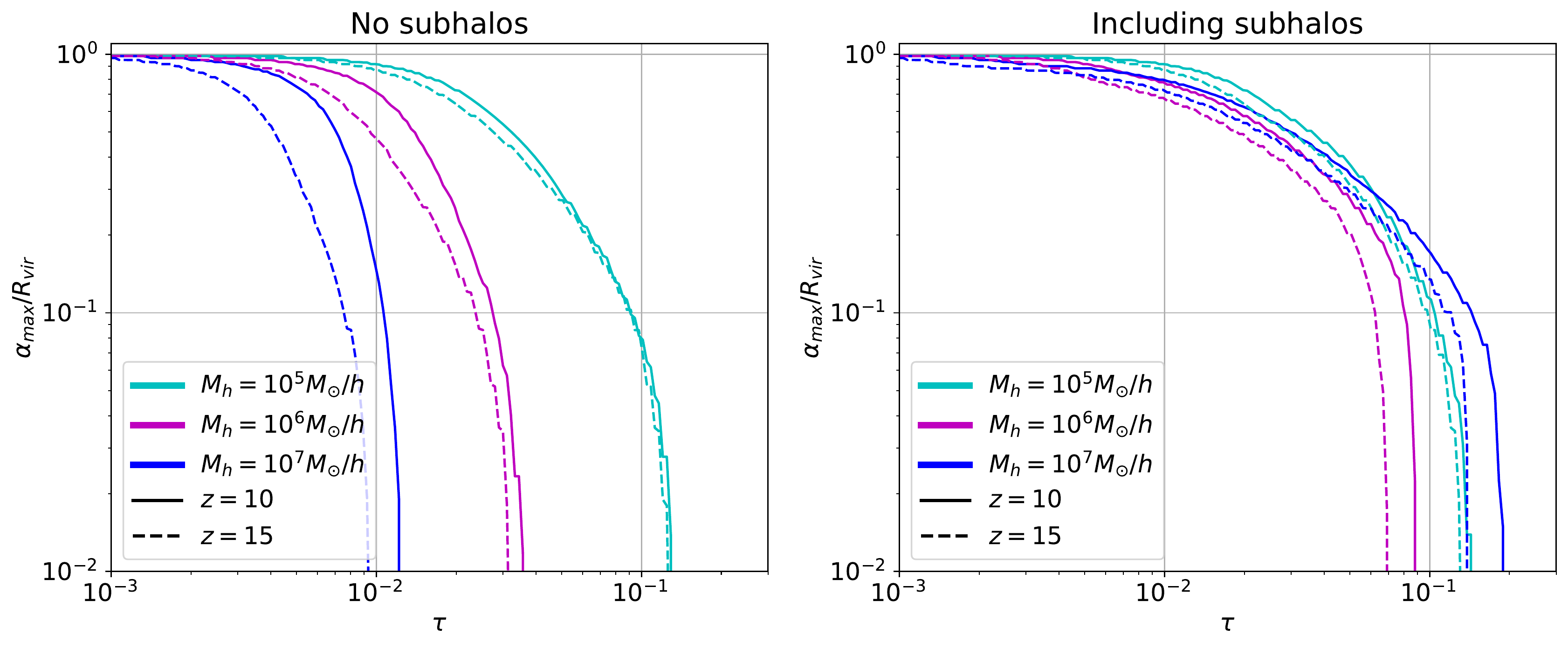}

    
   \caption{The maximum impact parameter $\alpha_{max}$ (normalized to the host halo virial radius) at which the optical depth exceeds a given $\tau$. Left: Only the host halo contribution to the optical depth is included. Right: Both subhalo and host halo contributions to the optical depth are included. The maximum impact parameter exceeding the required $\tau$ is enhanced due to the subhalos. 
   }
   \label{fig:impact1}
\end{figure}
\begin{figure}[t]

     \begin{tabular}{cc}
       \includegraphics[width=0.45 \textwidth]{
         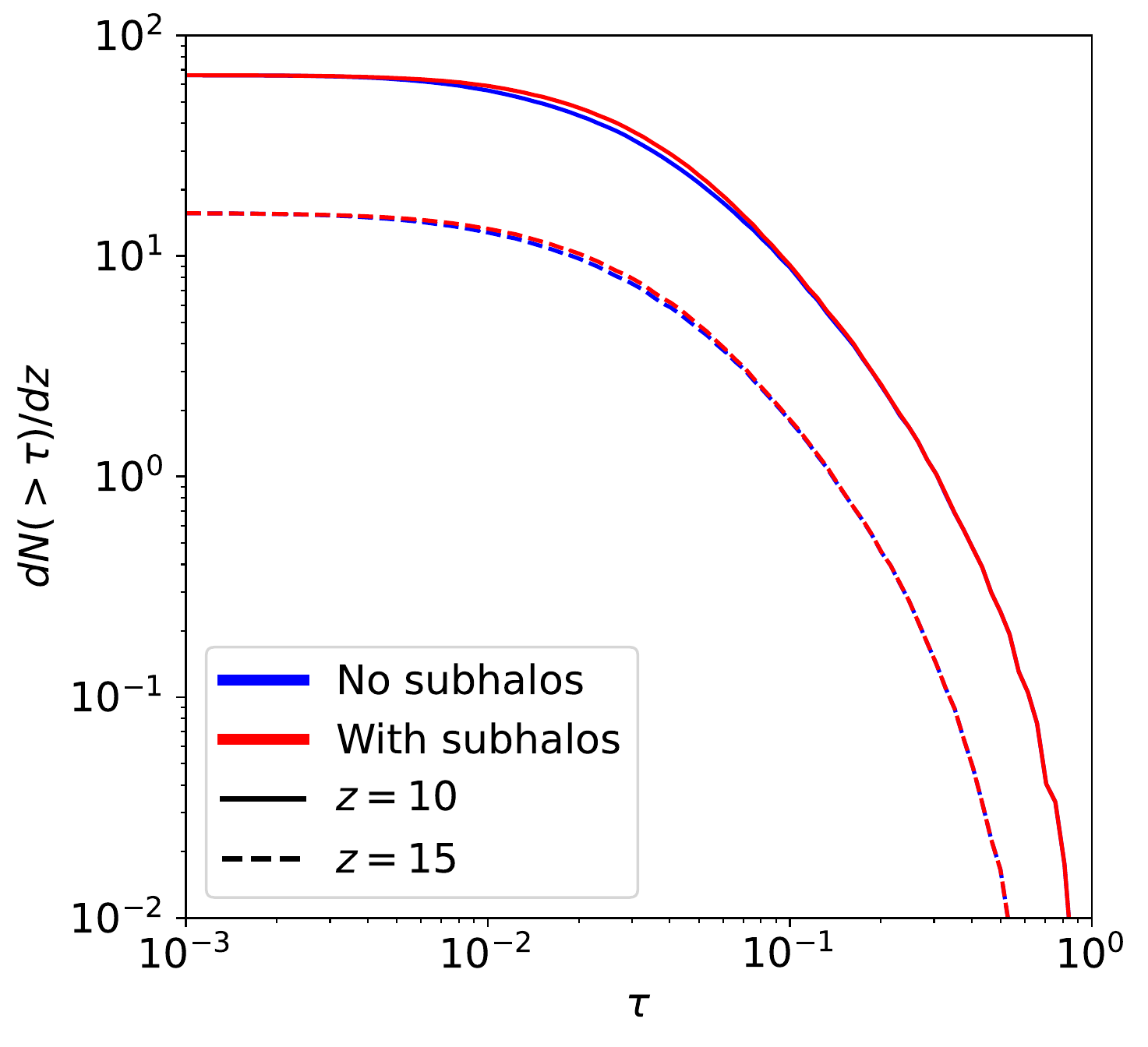}
       &
      \includegraphics[width=0.55\textwidth] {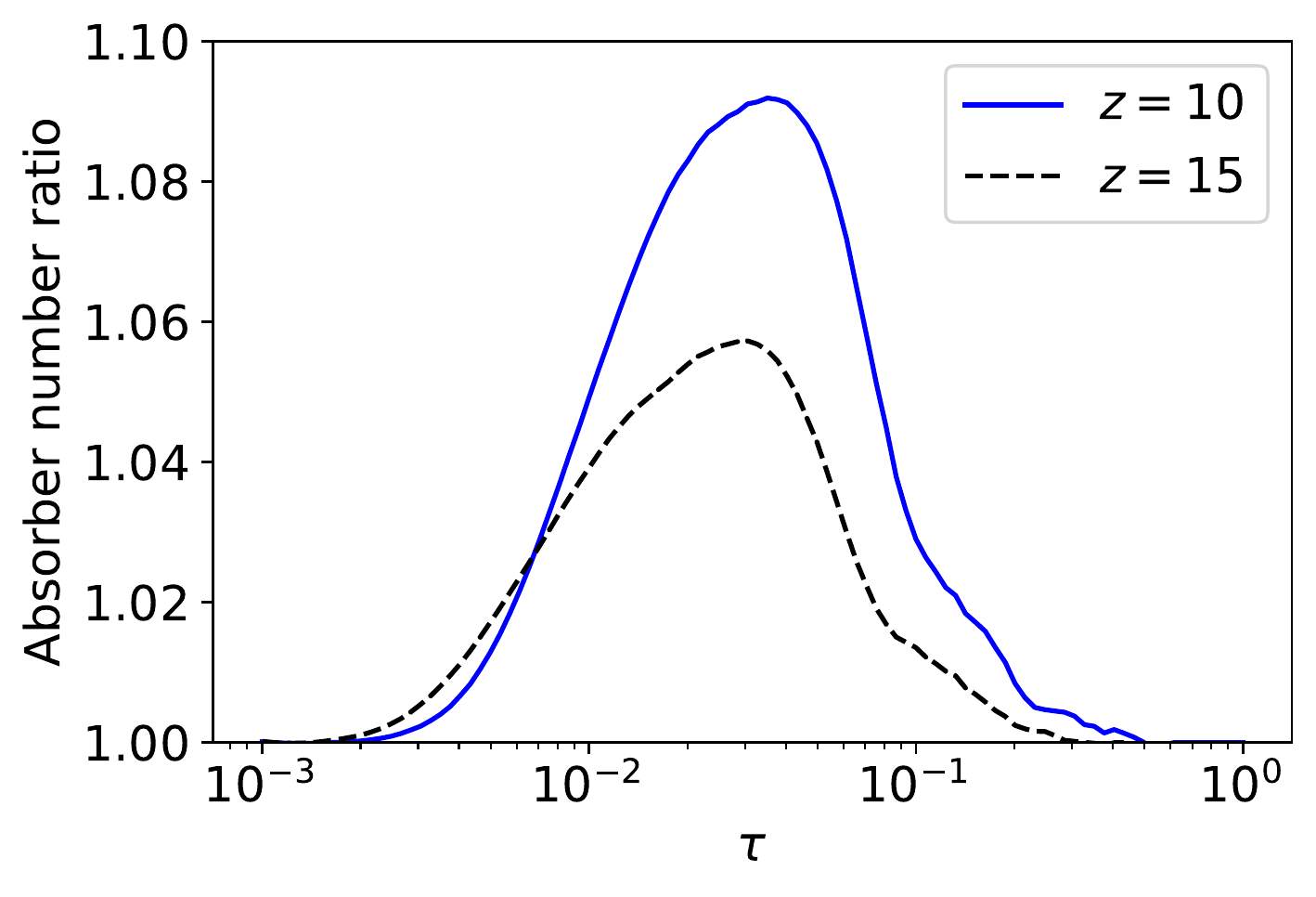}     
    \end{tabular}

    
   \caption{Left: The cumulative number of absorbers per redshift $dN(>\tau)/dz$ obtained by requiring the optical depth to exceed the observation threshold $\tau$. 'No subhalo' includes the contribution only from the host halos. 'With subhalo' includes the total contribution both from the host halos and subhalos. The subhalo enhancement is at most of order $10 \%$ and hard to see in this logarithmic axis. The ratio on the linear scale is shown in the right panel for a better inspection on the subhalo enhancement. Right: The ratio of the cumulative number of absorbers per redshift $dN(>\tau)/dz$ including the subhalo contribution to that without subhalo contribution. 
   }
   \label{fig:dNdtau}
\end{figure}
The maximum impact paramter at which the optical depth exceeds a given value of $\tau$ for a given host halo is plotted in the left panel of Fig. \ref{fig:impact1} as a function of $\tau$. For a large threshold value of $\tau$, the smaller host halo benefits from the smaller spin temperature and the effective geometric cross section ($\propto \alpha_{max}^2$) for a smaller halo can be bigger. For a small threshold value of $\tau$, on the other hand, the larger host halo benefits from the larger column density and its effective cross section can be bigger. The maximum impact parameter becomes bigger thanks to the subhalo contribution as illustrated in the right panel of Fig. \ref{fig:impact1} to be compared with the left panel which does not include the subhalo contributions.  The boost factor is bigger for a smaller impact paramter, and the subhalo contribution effect is more prominent for a smaller value of $\alpha_{max}$. The effect is bigger for a larger host halo as expected which tends to be more boosted by the subhalos than a smaller host halo. The subhalo effects are smaller for the smaller host halos because they have less subhalos and a smaller spin temperature.

The cumulative number of absorbers exceeding a given threshold optical depth $dN(>\tau)/dz$ is plotted in Fig. \ref{fig:dNdtau}. The curve labeled ``No subhalo" represents the abundance of 21 cm absorbers due to the minihalos without considering the subhalo contributions. The larger signals are expected at a lower redshift since more structures are formed at later times. The curve labeled as "With subhalos'' represents the abundance including both the host halo and the subhalo contributions. We can estimate the absorption line abundance in an analogous manner to the host halo calculation (Eq. \ref{dndz}) by considering the effective geometric cross section covering the region satisfying $ \tau^{total} = \tau^{h}(1+B)  >\tau$ for a given threshold value $\tau$. For instance, for the required optical depth of $3\times 10^{-2}$, the absorption line abundance increases by about 9 \% at $z\sim 10$ due to the subhalos compared with that including the host halo contribution alone.

Some comments regarding our simplified assumptions to model the subhalo gas properties are in order.
In addition to the lack of observational data, it is challenging to perform the desired numerical simulations because of the required high resolutions covering a wide dynamical range. For instance, even though we adopted the concentration mass relation for the dark matter profile from Ref. \cite{Ishiyama:2020vao} because it covers a wide range of mass and redshift for both host halos and subhalos, the subhalos tend to be more concentrated than the host halos in the numerical simulations \cite{Bullock:1999he}. Hence our estimation can be conservative and the actual optical depth may be bigger than our simple estimation. The cause of the higher concentration for the subhalos can be for instance the denser environment at their earlier formation epoch \cite{Moline:2016pbm, Giocoli:2007uv,Gao:2004au}. 

\begin{figure}[t]

     \begin{tabular}{ccc}
       \includegraphics[width=0.33\textwidth]{
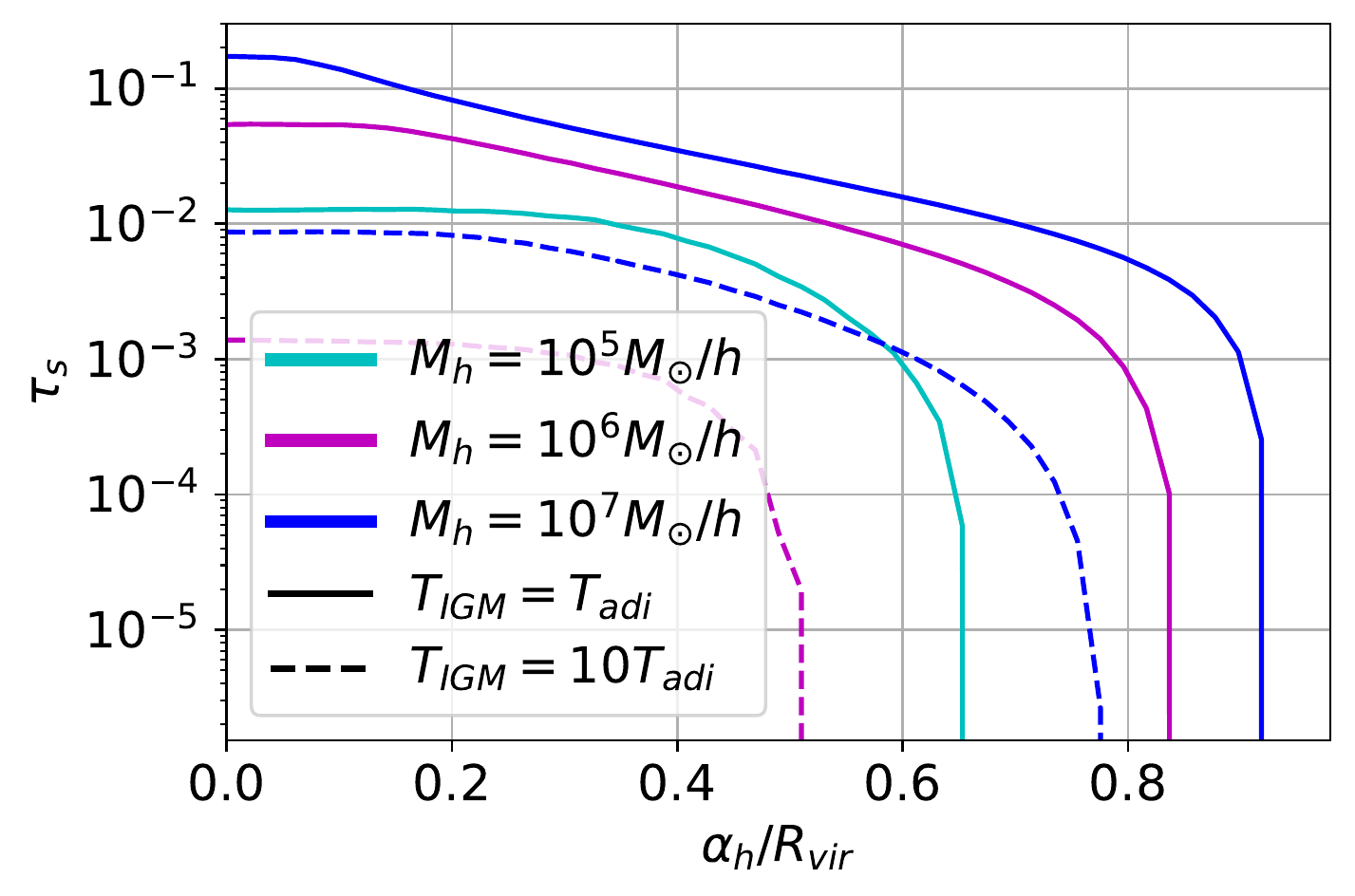}
       &
              \includegraphics[width=0.33\textwidth]{
               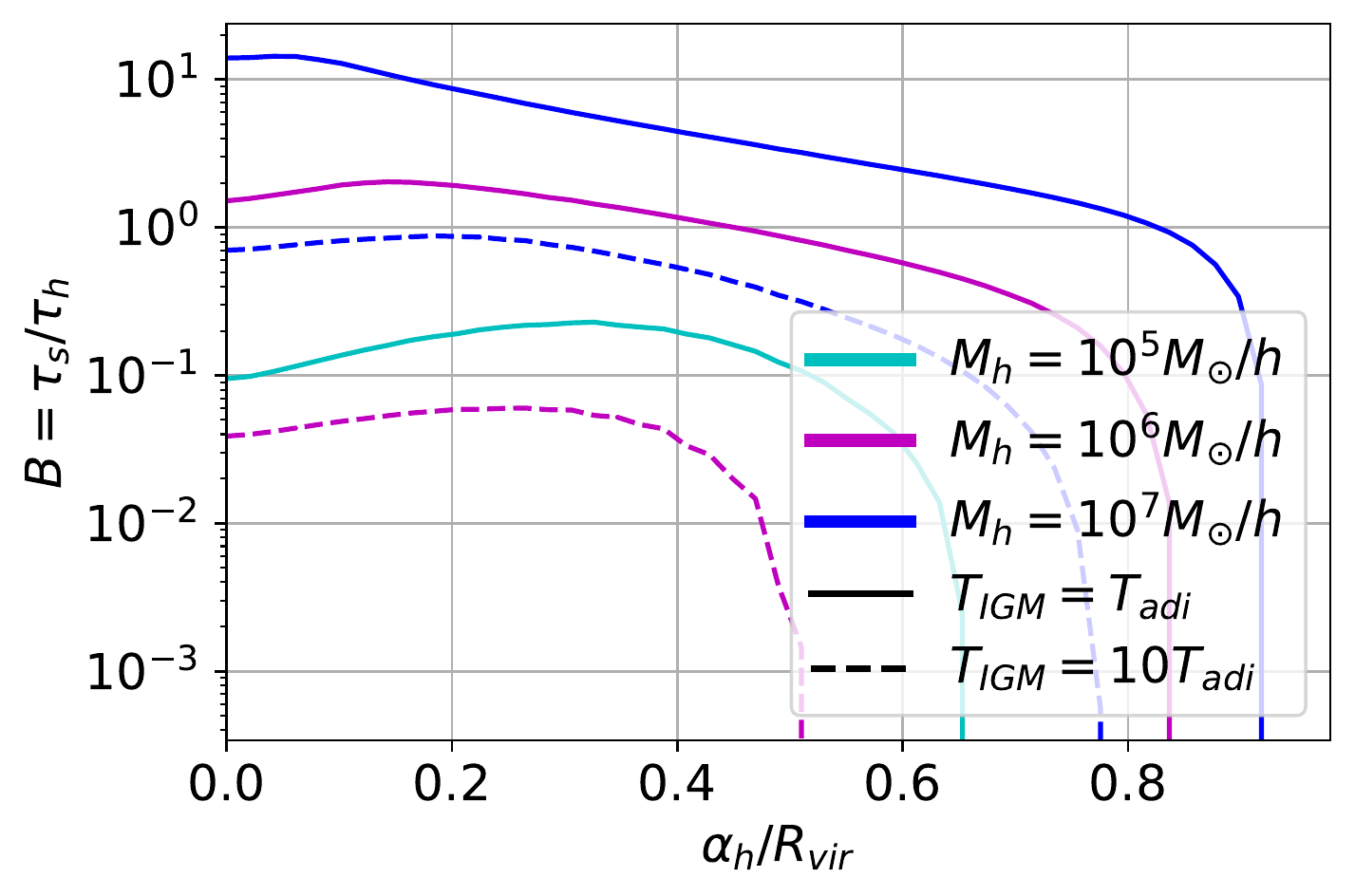}
              &
       \includegraphics[width=0.33\textwidth]{
         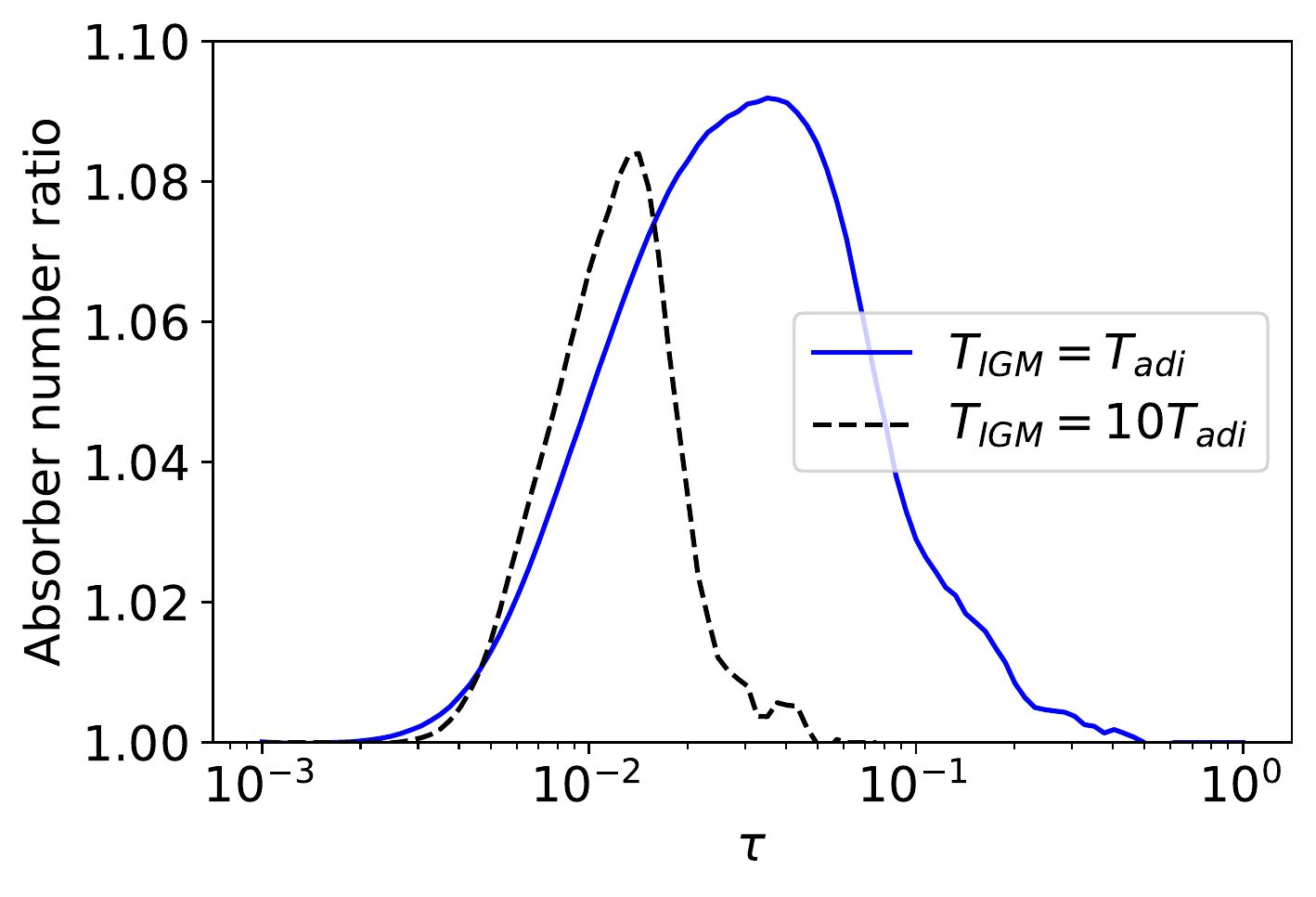}
    \end{tabular}

    
     \caption{The subhalo contribution to the optical depth at $z=10$ for the different IGM temperature. Left: The comparison of the subhalo optical depth at $z=10$ using $T_{IGM}=T_{adi}$ and $10 T_{adi}$ as a function of the impact parameter. The subhalo contribution for the host halo mass $M_{h}=10^5 M_{\odot}$ and $T_{IGM}=10 T_{adi}$ is negligible and absent in this figure. 
Middle: The corresponding boost factor. 
Right: The ratio of the cumulative number of absorbers per redshift $dN(>\tau)/dz$ including the subhalo contribution to that without subhalo contribution. 
}
   \label{fig:10Tadi}
\end{figure}

  On the other hand, there are effects which should reduce our simple estimation for the 21 cm signals. For instance, while we have used the adiabatically cooled IGM temperature in our discussions so far \cite{2001PhR...349..125B,2003tsra.symp...31M}, the IGM temperature is known to affect the 21 cm forest signal estimation because it can can change the lower mass limit of the relevant halos for the 21 cm signals \cite{2002ApJ...579....1F,2006MNRAS.370.1867F,1998MNRAS.296...44G}. We for simply used the Jeans mass for the lower mass limit which is proportional to $T_{IGM}^{3/2}$. The purpose of this paper is to estimate the ratio (rather than each magnitude) of the host halo and subhalo contributions, and the dependence on the IGM temperature may be somewhat relaxed in taking the ratio. We for illustration show the plots for the temperature which is 10 times the adiabatic case $T_{IGM}=10 T_{adi}$ in Fig. \ref{fig:10Tadi}. The left and middle panels show the subhalo optical depth and corresponding boost factor. The subhalo mass is smaller than the host halo mass, and consequently the range of the mass integration is more limited than the host halo mass range relevant to the 21 cm forest signals. For instance, the Jean mass is of order $10^5 M_{\odot}$ at z=10 if we assume $T_{IGM}=10 T_{adi}$ and the subhalo contribution to the optical depth is negligible for the host halo mass $M_{h}=10^5 M_{\odot}$. 
  The host halo optical depth as a function of the impact parameter for a given host halo is not affected by the change of $T_{IGM}$ and it is shown in Fig. \ref{fig:tau} (the IGM temperature affects our calculations when we integrate over the relevant halo mass function). 
  The effect of the IGM temperature on the number of 21 cm absorption lines is illustrated in the right panel of Fig. \ref{fig:10Tadi} which shows the ratio of the abundance between the total (host and subhalo) contributions and host halo contribution alone. The higher IGM temperature reduces the signals as expected by increasing the lower mass limit of the relevant halo mass. The larger optical depth part of this plot is affected more by removing the small halos in our calculations because a smaller halo has a smaller spin temperature and hence a bigger optical depth. It is relatively easy to realize the small value of $\tau$ even if the integrated mass range is reduced and a small $\tau$ region in this figure is not so affected. The IGM temperature evolution at the pre-ionization epoch is not well known and heavily model dependent, and we defer more detailed discussions on the influence of IGM temperature evolution on the subhalo gas properties and its consequence on the 21 cm forest signals to our forthcoming paper with the numerical simulations.
  
\begin{figure}[t]

     \begin{tabular}{ccc}
       \includegraphics[width=0.33\textwidth]
                       {
            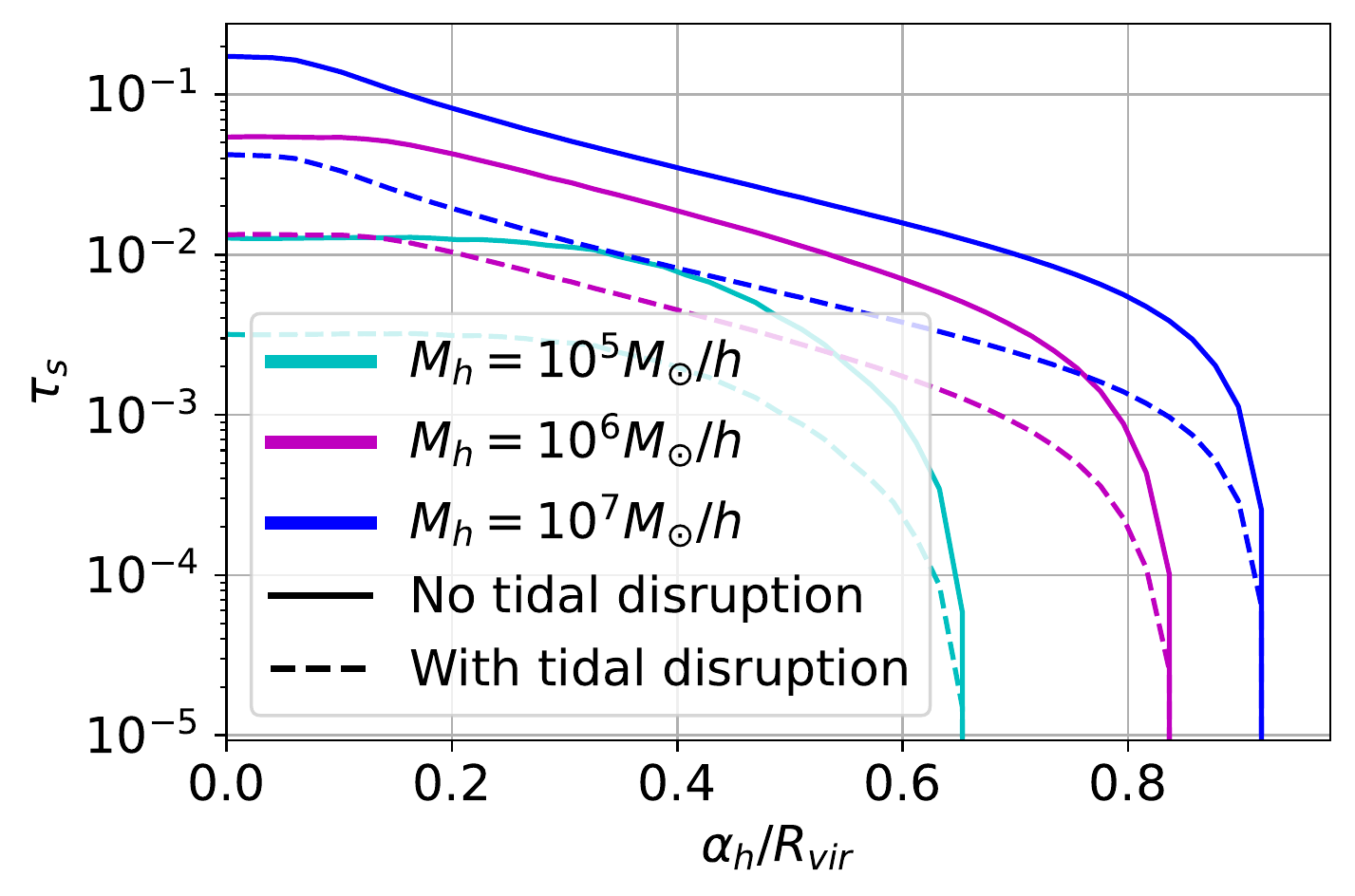}        
       &
                       \includegraphics[width=0.33\textwidth]{
                         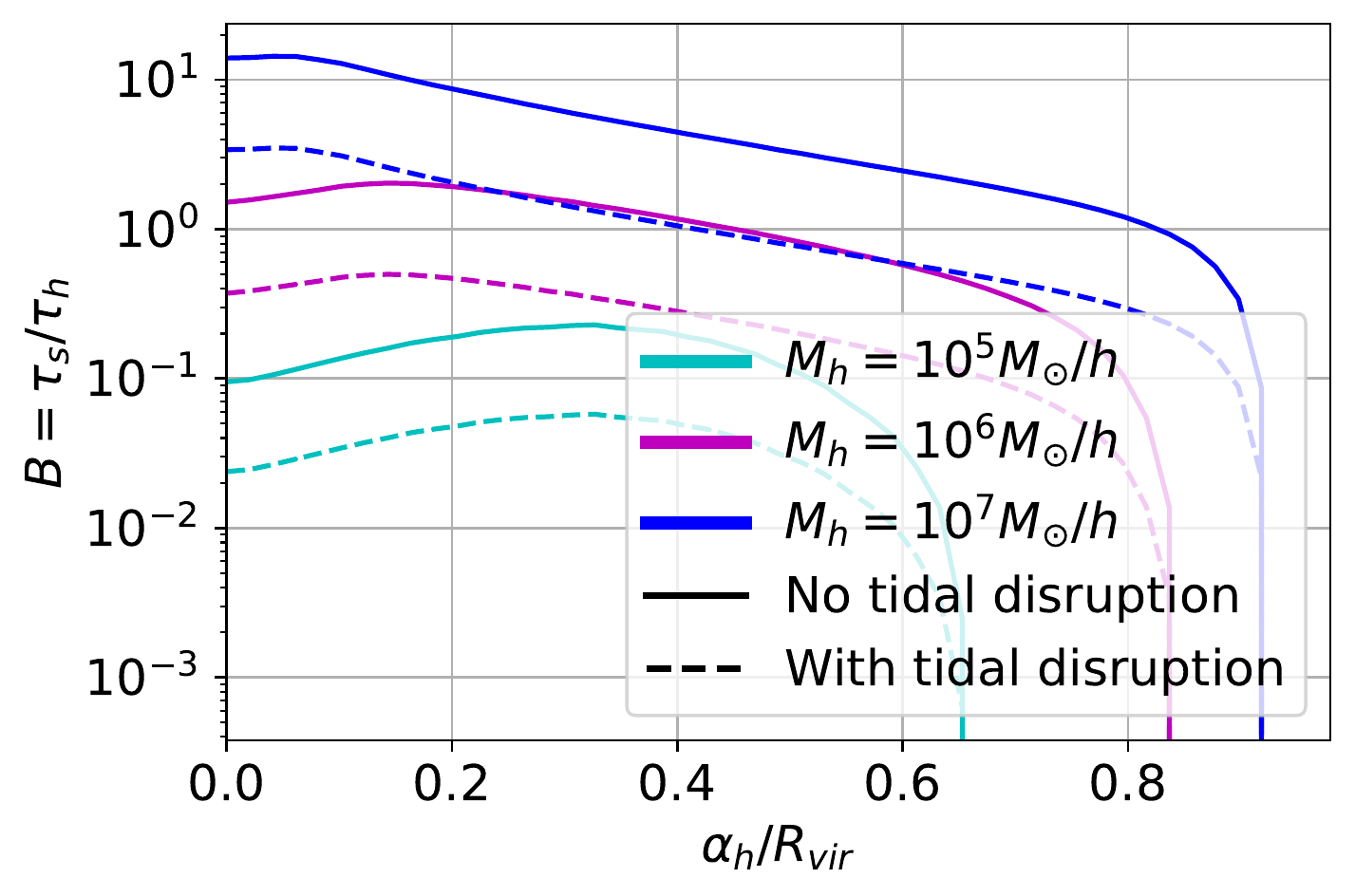}
                       &
                                     \includegraphics[width=0.33\textwidth]{
                                     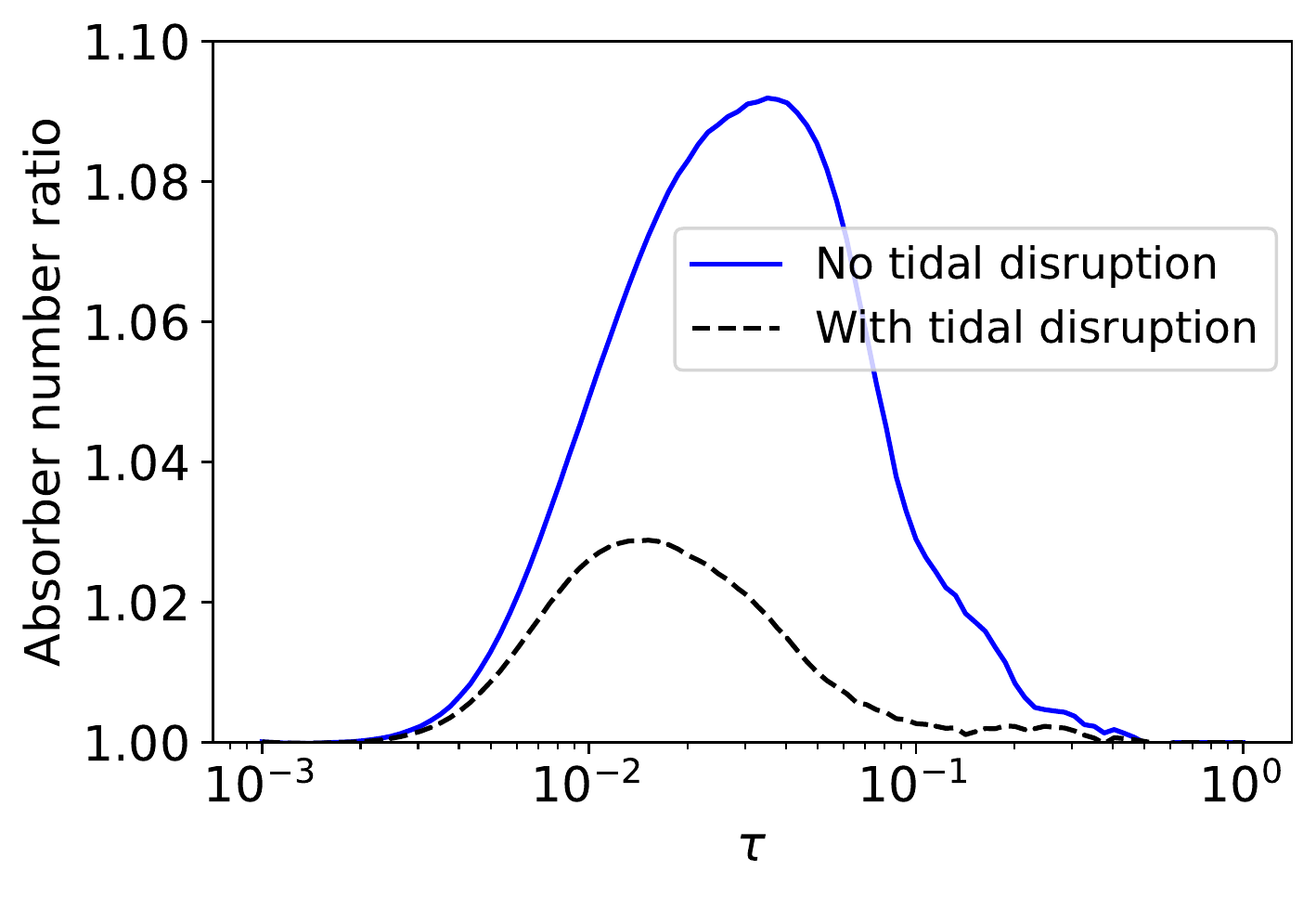}
    \end{tabular}

    
     \caption{The subhalo contribution to the optical depth at $z=10$ assuming the outer part of the subhalo is removed. Left: The subhalo contribution to the optical depth $\tau_s$ as a function of the impact parameter by removing the outer part of a subhalo beyond $0.77 r_s^{sub}$ ($r_s^{sub}$ is the scale radius of the subhalo). Middle: The corresponding boost factor. The host halo contribution to the boost factor $\tau_h$ (without including the subhalo contribution) is given in the left panel of Fig. \ref{fig:tau}. Right: The ratio of the cumulative number of absorbers per redshift $dN(>\tau)/dz$ including the subhalo contribution with respect to that without subhalo contribution.
   }
   \label{tau:tidal}
\end{figure}
\begin{figure}

     \begin{tabular}{ccc}
       \includegraphics[width=0.33\textwidth]{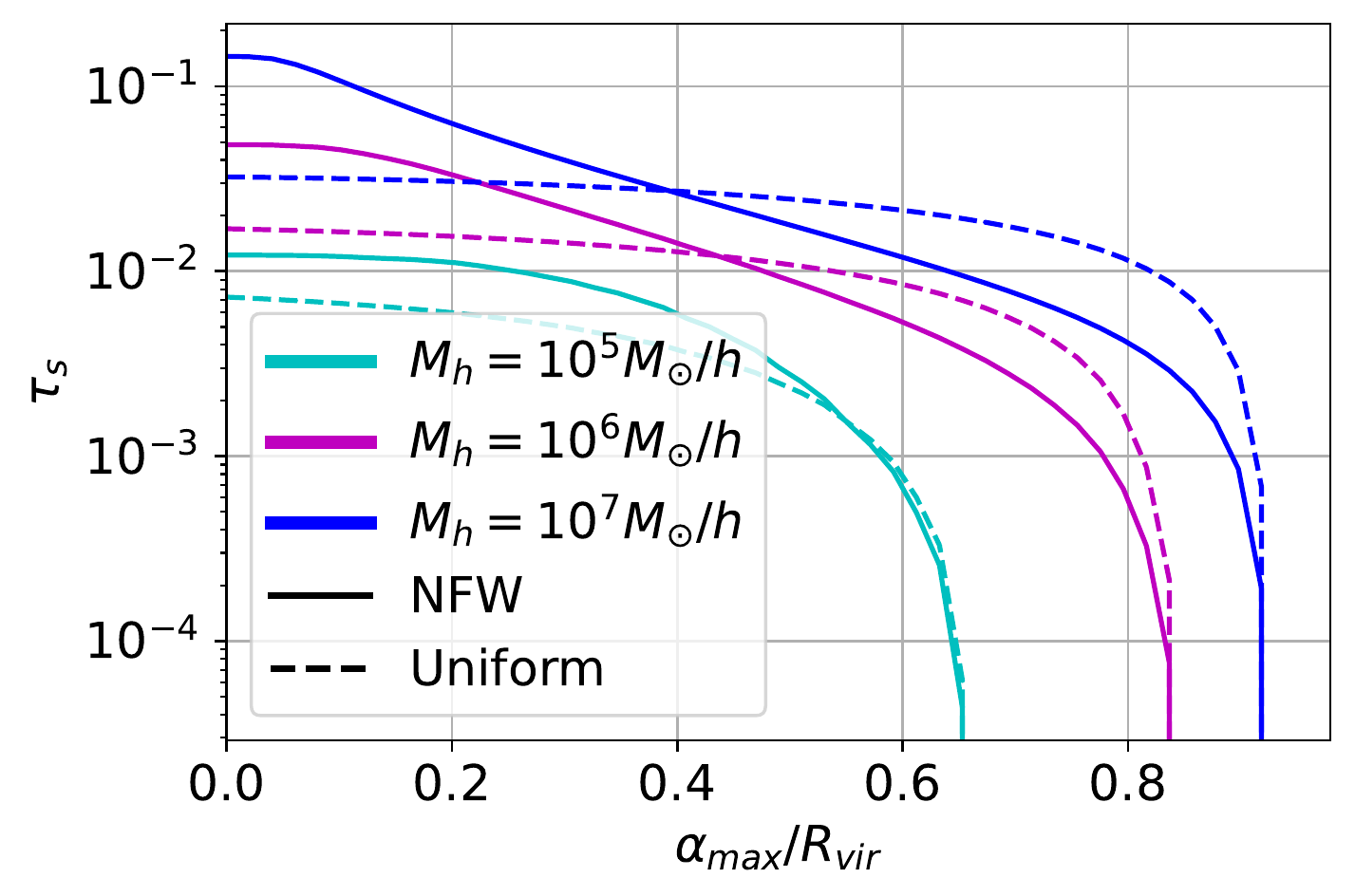}

&
       \includegraphics[width=0.33\textwidth]{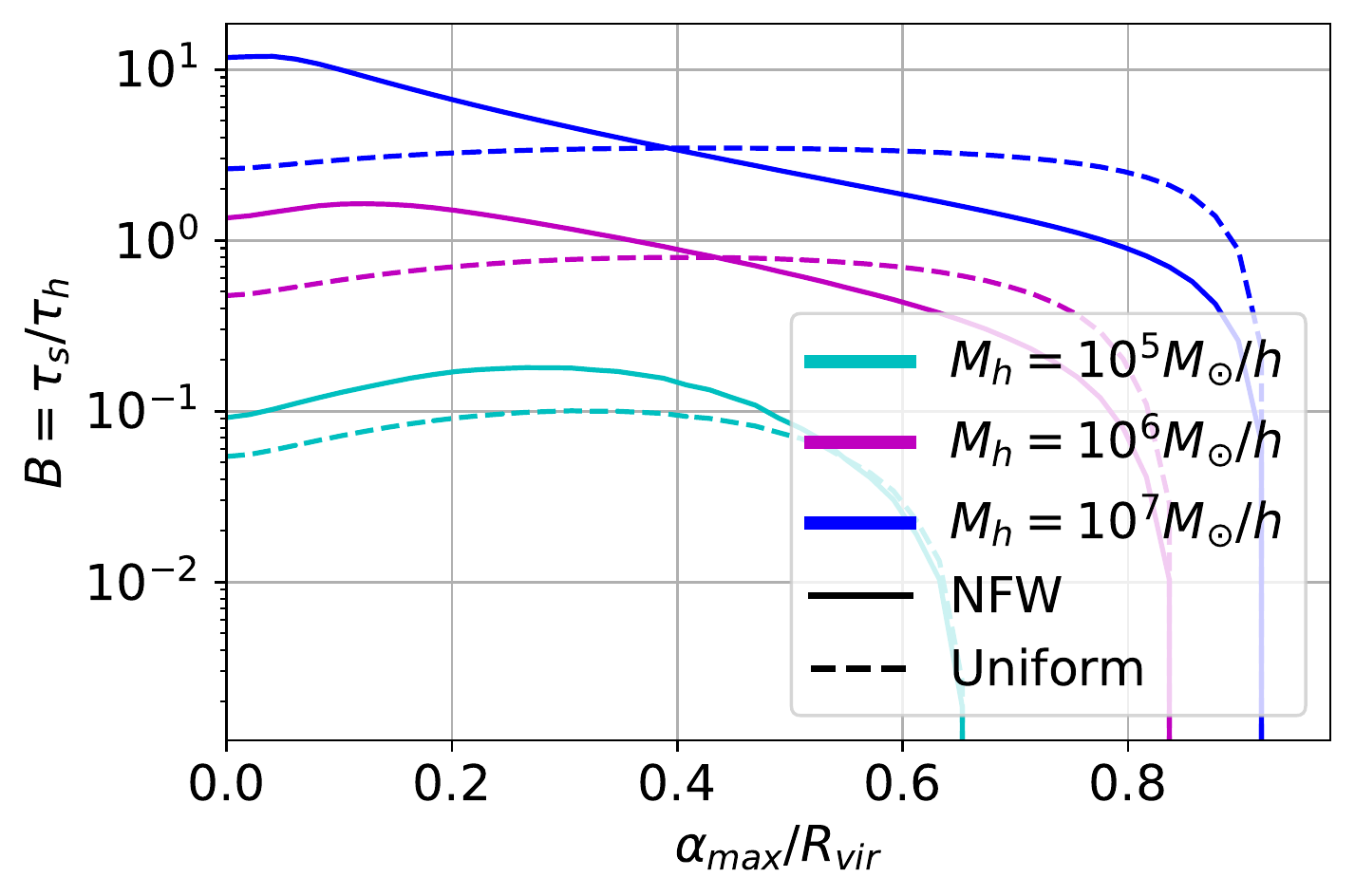}

       \includegraphics[width=0.33\textwidth]{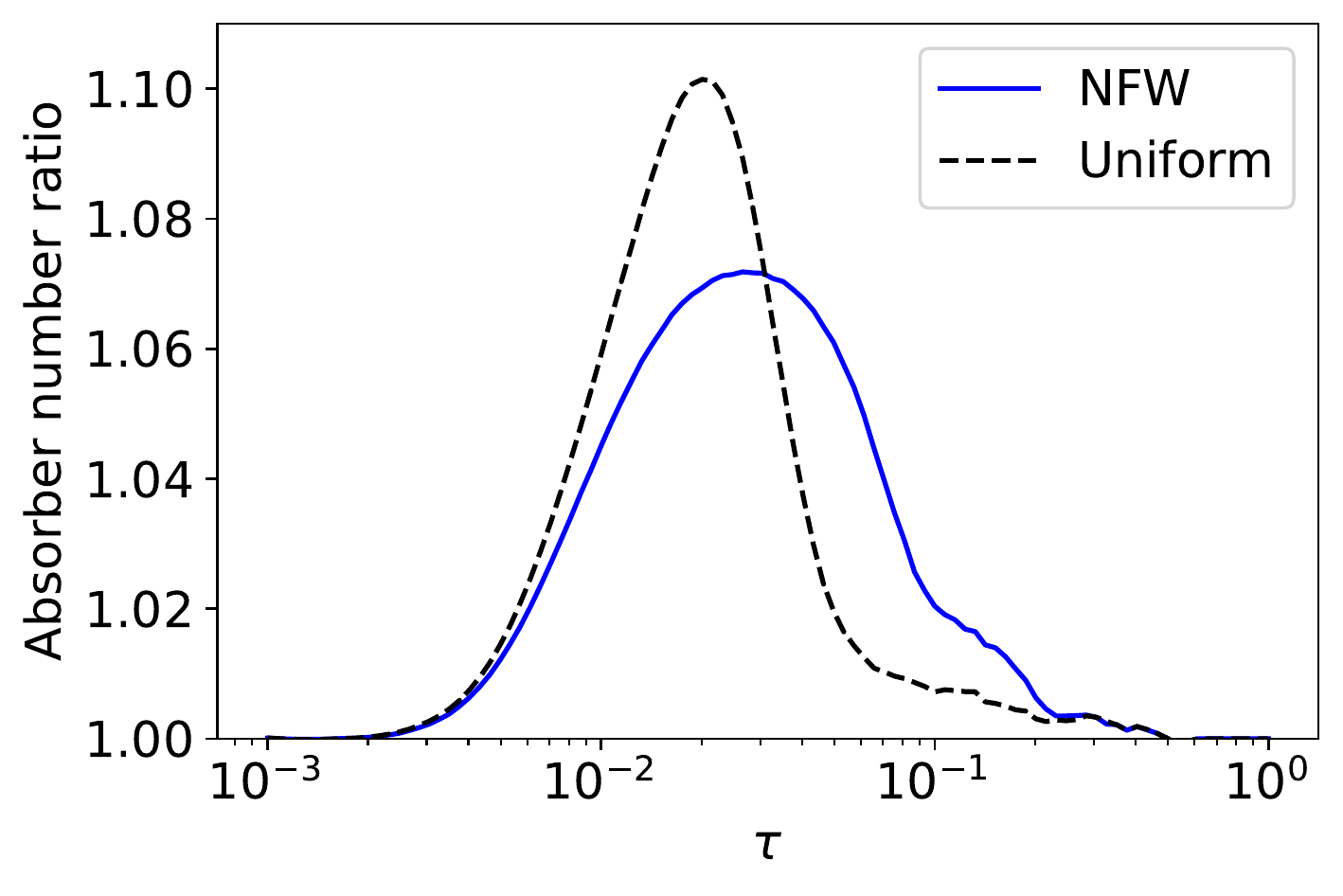}
          
                 \end{tabular}

    
     \caption{The subhalo contribution to the optical depth at $z=10$ for the different subhalo spatial distribution inside a host halo: uniform and NFW profiles. Left: The subhalo contribution to the optical depth $\tau_s$ (without including the host halo contribution) as a function of the impact parameter for the NFW spatial distribution and constant uniform distribution. Middle: The corresponding boost factor. The host halo contribution to the boost factor $\tau_h$ (without including the subhalo contribution) is given in the left panel of Fig. \ref{fig:tau}. Right: The ratio of the cumulative number of absorbers per redshift $dN(>\tau)/dz$ including the subhalo contribution to that without subhalo contribution.
   }
   \label{fig:tauprofile}
\end{figure}
Another suppression of the estimated 21 cm signal may come from the reduction of the gas bounded to a subhalo. For instance, while the subhalos orbit in their host halo, their gas may be stripped away due to the tidal forces and/or the ram pressure.
For illustration, we removed the gas for the outer region $r>0.77 r^{sub}_s$ ($r_s^{sub}$ is the subhalo scale radius) in the subhalos \cite{Hayashi:2002qv} and the subhalo contribution to the optical depth is shown in Fig. \ref{tau:tidal}. While the significant part of 21 cm forest signals can come from the dense inner core of a halo, the outer part of a halo can also contribute because of its larger volume and the lower temperature. Removing the gas content of subhalo outer region can give appreciable effects. The expected reduction in the absorber abundance is shown in the right panel of Fig. \ref{tau:tidal}.
It is still under an active debate if the dark matter and gas can remain in the tightly bound core after they are stripped away in the outer parts of a subhalo \cite{Spinelli:2019smg, Villaescusa-Navarro:2018vsg,Han:2015pua,Springel:2008cc,Stref:2016uzb,Kelley:2018pdy,Hayashi:2002qv}. For instance, while there have been claims that the subhalos whose tidal radii are less than $0.77 r_s^{sub}$ are completely disrupted, the complete loss of subhalo identity may well be the numerical artifact \cite{Hayashi:2002qv,vandenBosch:2017ynq}. 
Our study focuses on the pre-reionization epoch $z\gtrsim 10$ (corresponding to the age of the Universe smaller than 0.5 Gyr), and, compared with the post-ionization observations of the subhalos, the tidal effects can be less severe because of the shorter time during which the subhalos are subject to the tidal disruptions. The neutral gas reduction due to the ionization can also be less severe because of the lack of ionization sources at a high redshift of our interest. The numerical studies would be required to clarify those issues on the subhalo gas properties which we plan to explore in our forthcoming paper with more detailed numerical simulations.

Another non-negligible effects also may come from the spatial distribution of subhalos in a given host halo.
Even if the subhalos initially follow the underlying host halo dark matter distribution (we adopted the NFW distribution in our setup so far), the subhalo tends to be distributed more in the outer region of the host halo under the influence of the tidal disruptions \cite{Han:2015pua,Springel:2008cc,Stref:2016uzb,Kelley:2018pdy,Hayashi:2002qv,Moline:2016pbm}. For illustration, we also estimated the signals adopting the uniform constant distribution which spreads the subhalos more to the outer part compared with the NFW distribution (the corresponding density distributions are in Eqs. \ref{eq:weight1}, \ref{eq:weight2}). Fig. \ref{fig:tauprofile} shows the corresponding optical depth and the resultant absorption line abundance. A given host halo has a bigger optical depth at a smaller impact parameter. When the subhalos are more concentrated in the host halo center, the optical depth can become even bigger. Hence, when the minimum required optical depth is bigger, the absorption line abundance enhancement can be bigger due to the subhalos for the NFW distribution which is more centrally concentrated than the uniform distribution. On the other hand, when the host halo can exceed the required minimum optical depth at the small impact parameter even without the subhalos, it is more advantageous to have more subhalos at an outer region where the host halo alone cannot give a big enough optical depth. This is reflected in the bigger absorption line abundance at a smaller $\tau$ for the constant density distribution than the NFW distribution.

\section{Discussion}
\label{sec:disc}
The boosting of the 21 cm optical depth is caused by the cold gas in subhalos. Therefore some hydrodynamic heating effects that we do not take into account may affect our results. One of the possibilities is the thermal conduction. We can estimate the typical time scale for the thermal conduction from the host halo to the subhalo as $ T_{vir}/(dT/dt)$, where $T_{vir}$ is the subhalo virial temperature and the temperature variation time scale $dT/dt$ can be estimated from the heat conduction equation \cite{1953ApJ...117..431P}. For instance, using the host halo mass of $M_h=10^7 M_{\odot}, m_s=10^5 M_{\odot}$ at $z=10$ along with the thermal conductivity in Ref. \cite{1953ApJ...117..431P}, the heating timescale is of order $\sim1{\rm Gyr}$ which is $\sim100$ times longer than the dynamical time of the system (free-fall time scale of the subhalo). Thus heating due to the thermal conduction can be neglected. This is not surprising considering that we demand the lack of ionization and consequently the lack of the ionizaed electrons which could work as the efficient thermal conductor (note the mean free path for heavier elements are suppressed). 
Another possibility is the compressional heating. In the outer low density region of a host halo, the compressional heating is unlikely effective because the pressure in this region is expected to be lower than that in a subhalo. 
In the central high density region, on the other hand, not only the compressinal heating but also the tidal stripping would work effectively. We will confirm what happens in such a complicated situation by numerical hydrodynamic simulations in the forthcoming paper.


Our simple analytical estimation could demonstrate the potential significance of the subhalos in the 21 cm forest signals, and the 21 cm forest observation may well serve as an intriguing probe on the gas content of the substructures at the pre-reionization epoch. Before concluding our discussions, let us discuss the future prospects for the detectability of the 21 cm forest. Following Refs. \cite{2006MNRAS.370.1867F,2002ApJ...579....1F,2021MNRAS.506.5818S, 2015aska.confE...6C,Villanueva-Domingo:2021cgh}, using the radiometer equation, the required minimum flux density $S_{min}$ of a radio background source for the 21 cm forest observation reads
\ba
S_{min}-S_{dam}= \frac{2 k_B T_{sys} A_{eff} } { \sqrt{\Delta \nu t_{int}} } S/N
\ea
where $S_{dam}=S_{min}e^{-\tau}$ is the damped flux representing the absorption features, $\tau$ is the 21 cm optical depth, $S/N$ is the signal to noise ratio, $\Delta \nu$ is the bandwidth, $A_{{\rm eff}}/T_{{\rm sys}}$ is the ratio between an effective collecting area and a system temperature and $t_{{\rm int}}$ is the integrated observation time.
We can then estimate the required minimum flux from the source for a given signal to noise ratio as
\begin{equation}
\begin{split}
  S_{\rm min}=10.4\rm{mJy}\left(\frac{0.01}{\tau}\right)\left(\frac{S/N}{5}\right)\left(\frac{\rm{1 kHz}}{\Delta \nu}\right)^{1/2}
 \left(\frac{5000[\rm m^{2}/K]}{A_{\rm{eff}}/T_{\rm{sys}}}\right)
  \left(\frac{100~\rm{hr}}{t_{{\rm int}}}\right)^{1/2}.
\end{split}
\label{eq:Smin}
\end{equation}
The numerical values adopted in Eq. \ref{eq:Smin} correspond to those for the SKA-like specifications. 
While the 21 cm forest has the great advantage for its being free from the diffuse foregrounds by looking at bright sources for a long time, its observations rely on the existence of the radio-loud sources at a high redshift ($z\gtrsim 6$). The simple estimation given by Eq. \ref{eq:Smin} indicates the necessity of bright high redshift sources with a flux of order 10 mJy. The simple extrapolation of observed radio loud sources to a higher redshift gives us an estimate of the order ${\cal O}(10^4)$ radio sources at $z \sim 10$ bright enough to be detected \cite{Haiman_2004,Xu:2010us,Ivezic:2002gh}, and hundreds of bright radio sources at a $z\gtrsim 6$ redshift are projected to be detected in the coming years using the on-going radio surveys such as the LOFAR Two-metre Sky Survey (LoTSS) \cite{2017A&A...598A.104S} and Giant Metrewave Radio Telescope (GMRT) \cite{1991ASPC...19..376S}.
Even though the existence and detection of such sources are not guaranteed, it is encouraging that so far about 10 radio-loud sources around $z\sim 6$ have been detected \cite{Spingola:2020iaq,2020NatCo..11..143A, ban2015,ban2018,Banados:2021imw,Zhang:2022dpm,2021MNRAS.506.5818S,2021A&A...648A...3K, 2021A&A...647L..11I} (e.g. the blazer at $z\sim 6.1$ which was observed by NVSS (NRAO VLA Sky Survey) with $\sim 20$ mJy flux at 1.5 GHz (and also measured by other observations at another frequency including the GMRT with $\sim 60$ mJy flux at 147 MHz) \cite{Spingola:2020iaq, 2017A&A...598A..78I}.)
The future prospects of the 21 cm forest are hence promising considering the potential existence of the high redshift bright sources. The gas content of the subhalos can be an interesting science topic to explore for the future 21 cm forest observations.

While we plan to present the detailed numerical studies in our forthcoming paper, let us briefly mention the issues regarding the survival of the low-temperature gas in the subhalos \footnote{We thank the referee for urging us to comment on this point.}.
We have investigated the effects of the ram pressure by performing SPH simulations to be presented in the separate forth-coming paper. In our pilot simulations, we, for instance, have checked the evolution of a subhalo with its mass $10^5M_{\odot}$ initially moving at the virial radius of a hosthalo with $10^7 M_{\odot}$ and have found the followings;
(i) Under the impact of the ram pressure, the gas temperature in the subhalo increases up to $\sim 300K$ (about the twice of the initial temperature of the subhalo when it starts infalling into the host halo). Since the temperature in the subhalo is still much lower than the temperature of the host halo, the merging of the subhalos into the host halo does not affect our qualitative discussions on the subhalo's contributions to boost 21-cm optical depth.
(ii) Even if the ram pressure does not strip the gas core from the subhalo, it can reduce the angular momentum of the subhalo. Such an effect can shorten the time scale for the infall into the core of the host halo. We however found the subhalos can still retain the enough gas for $\gtrsim 0.1-0.2 Gyrs$ inside the host halo (note the age of the Universe is $\sim 0.5 Gyrs$ at $z\sim 10$) even if they are eventually tidally destroyed at a lower redshift ($z<10$) for the parameter range of our interests discussed in our paper. 
We hence expect the qualitative pictures based on the analytical formulation in this paper still hold and will present the detailed numerical simulation results in our forthcoming paper.

We have shown the 21 cm optical depth can be enhanced due to the presence of subhalos within the minihalos. Because the larger host halo has a smaller optical depth and a larger subhalo boost factor, the subhalo contributions would be of particular interest for the large host halos whose optical depth is too small to be observed without the subhalos. We demonstrated that the optical depth for a large host halo (the mass of order $10^7 M_{\odot}$) can be boosted by an order of magnitude. While this boost effect is off-set by the smaller abundance for a bigger halo, the resultant absorption line abundance which is obtained by integrating over all the relevant halo mass range can be enhanced by up to 10\%.
Even though the actual effects may be reduced due to the large uncertainties such as the gas temperature and gas mass evolutions at the pre-ionization epoch, our analytical illustration for the boost factor can be the first step for the more detailed study including the dedicated numerical simulations to clarify the HI distribution in the substructures for the potential enhancement of the 21 cm forest signals.
We plan to numerically investigate some of those issues on the subhalo gas properties such as the subhalo gas profile evolution in the host halo in a forthcoming paper to give more realistic estimation for the 21 cm forest signals. 
   
This work in part was supported by the JSPS grant numbers 18K03616, 17H01110, JST AIP Acceleration Research Grant JP20317829 and JST FOREST Program JPMJFR20352935.

\appendix
\section{Subhalo mass function}
We adopted the subhalo mass function $dn_s/dm_s$ for a given host halo mass $M_h$ discussed in Ref. \cite{Ando:2019xlm}
\ba
m_s^2 \frac{dn_s}{dm_s}
=
(a+b m_s^{\alpha}) \exp
\left[
  -\left(
\frac{m_s}{m_c}
  \right)^{\beta}
  \right]
\ea
where
\ba
a(M_{h},z)=6.0 \times 10^{-4}\times (1+z) M_{h}(z)
(2-0.08 \log_{10}[M_{h}(z)])^2
\left(
\log_{10}\left[
\frac{M_{h}(z)}{10^{-5}}
  \right]
\right)^{-0.1}
\ea
\ba
b(M_{h},z)=8.0 \times 10^{-5} M_{h}(z)
(2-0.08 \log_{10}[M_{h}(z)])
\\
\times
\left(
\log_{10}\left[
\frac{M_{h}(z)}{10^{-5}}
  \right]
\right)^{-0.08z}
\left(
\log_{10}\left[
\frac{M_{h}(z)}{10^{-8}}
  \right]
\right)^{-1}
\left(
\log_{10}\left[
\frac{M_{h}(z)}{10^{18}}
  \right]
\right)^{2}
\ea
\ba
m_c=0.05(1+z)M_{h}(z)
\\
\alpha=0.2+0.02z
\\
\beta=3
\ea

\end{document}